# Periodic recurrent waves of Covid-19 epidemics and vaccination campaign

Gaetano Campi[1,2], Antonio Bianconi[1,2]

[1] Institute of Crystallography, Consiglio Nazionale delle Ricerche CNR , via Salaria Km 29.300, Monterotondo Roma, I-00015, Italy;
[2] Rome International Centre Materials Science Superstripes RICMASS via dei Sabelli 119A, 00185 Rome, Italy

**Abstract**
While understanding of periodic recurrent waves of Covid-19 epidemics would aid to combat the pandemics, quantitative analysis of data over a two years period from the outbreak, is lacking. The complexity of Covid-19 recurrent waves is related with the concurrent role of i) the containment measures enforced to mitigate the epidemics spreading ii) the rate of viral gene mutations, and iii) the variable immune response of the host implemented by vaccination. This work focuses on the effect of massive vaccination and gene variants on the recurrent waves in a representative case of countries enforcing mitigation and vaccination strategy. The spreading rate is measured by the ratio between the reproductive number $R_t(t)$ and the doubling time $T_d(t)$ called *RIC-index* and the daily fatalities number. The dynamics of the Covid-19 epidemics has been studied by wavelet analysis and represented by a non-linear helicoid vortex in a 3D space where both *RIC-index* and fatalities change with time. The onset of periodic recurrent waves has been identified by the transition from convergent to divergent trajectories on the helicoid vortex. We report a main period of recurrent waves of 120 days and the elongation of this period after the vaccination campaign.

## 1. Introduction

Periodic recurrent waves occur in out of equilibrium dynamics for propagation into unstable states in viral epidemics [1-4], in population dynamics, ecology, cyclic predator-prey systems, spreading fires, ferroelectrics and membranes [5-20]. The viral pandemic Influenza A appearing in 1918 (called Spanish flu) has shown three main waves in its first two years [1] folowed by seasonal waves of influenza variants for many years [2]. The study of periodic waves in the case of measles epidemics has pointed out the role of spatial hierarchy of host population structure and the role of vaccination [3]. The complexity of recurrent waves in epidemics is driven by either i) the spatiotemporal heterogeneities and mobility of population constraining the virus diffusion described by network theory in the macroscopic physical world and ii) the gene point mutations, i.e., atomic substitutions and deletions affecting only one or few nucleotides in the viral gene sequence and in the immunity system of the host cells, which are made up of active atoms which seem to make up unpredictable choices in the microscopic quantum world.

The Covid-19 pandemic has infected 327 million people with 5.5 million fatalities [21-38] in two years from the outbreak, and it has been contrasted in different countries by two policies: i) the "zero Covid" [29-35] and the "mitigation" strategy [36]. In 2021 the vaccination campaign started and new variants appeared [39]. Mathematical models of Covid-19 spreading have been developed based on standard epidemiological theories [25-28] and on network theory [29-35], but quantitative analysis of recurrent waves and the effect of vaccination campaign on these recurrent waves is lacking. The SARS CoV-2 contagiousness is measured here by the ratio of the time dependent reproductive number, $R_t$ and the doubling time, $T_d$, called *RIC-index* [21]. In previous works [22-24] we have identified the *critical* doubling time $T_d^*$ which separates the explosive *supercritical* regime ($T_d < T_d^*$) from the arrested *subcritical* regime ($T_d > T_d^*$) of the epidemics spreading. The severity of the impact of the epidemics on the population has been quantified by the number of daily deaths per million population, *Df*. The experimental periods of the recurrent waves have been extracted by the Wavelet Transform (WT) [40,41] of either the *RIC-index* and the *Df* time series to extract the frequency (or period) of non-stationary time series data [40-42].

The periodic traveling waves of SARS-CoV-2 are represented in this work by a Covid-19 acentric helicoid vortex in a 3D space by combining the time evolution of daily deaths per million population, *Df*, and *RIC-index*.

To quantify the divergence/convergence of non-stationary orbits of the Covid-19 helicoid vortex we have used the Rosenstein algorithm [43] to calculate the Largest Lyapunov Exponents (LLE) of the Covid-19 to identify the onset of periodic recurrent waves. This approach allows monitoring discontinuities and fluctuations at small timescale in complex systems [43-48] characterized by short time series and intrinsic transient character where the conventional chaotic deterministic non-linear dynamical approach characterized by the classical Lyapunov exponents fails.

## 2. Results and Discussion

The Covid-19 spreading rate shows a heterogeneous character due to the different containment measures adopted in the different geographical areas. Here we focus on Italy as representative of the West European Area. Data on epidemic spreading and severity from 24/02/2020 to 15/12/2021 have been taken from the recognized public database *OurWorldInData* [49] to extract the time-dependent doubling time $T_d$ as described in [24] and the time-dependent reproductive number $R_t$ from [49]. The time evolution of both $T_d$ and $R_t$ is shown in **Figure 1a**, where the grey horizontal strip indicates the *critical* phase defined by $50<T_d^*<100$ days [22,24] and $1<R_t^*<1.1$ [28].

This critical strip separates the *supercritical* regime by the *subcritical* regime. In the *supercritical* regime $T_d$ grows up exponentially assuming values less than 100 days, while $R_t$ values result larger than 1. In the subcritical regime $T_d$ becomes large enough ($T_d>100$), $R_t$ becomes less than 1 and the exponential growth is arrested [21]. The explosive supercritical regime with exponential growth occurred at the threshold of the first two 2020 Covid waves, indicated by red areas in the panel (a), corresponding to a peak in $R_t$ joint with a dip of $T_d$, due to their anti-correlation. The green shadowed vertical rectangle indicates the time of the first lockdown with strict rules (SL) enforced by the "mitigation" policy in the first wave (from 24/02/2020 to 30/04/2020) due to the first variant from Wuhan provided by reference [37]. The red shadowed rectangle indicates the time of the mild lockdown with loose rules (ML) to face the European *E20* variant diffusion in the second wave (from 25/10/2020 to 31/12/2020) getting out of the *supercritical* phase. A third small pocket with critical conditions, occurs around day 430, due to the *alpha* variant diffusion contrasted by the start of the vaccination campaign at $t_v=400$ day. The massive vaccination campaign in Italy has been able to freeze the outbreak of the *delta* variant. In fact, while $R_t$ shows a sharp peak, reaching values larger than 1 but the doubling time $T_d$ shows a minimum that remains much longer than 100 days, therefore the epidemic spreading has been frozen in the *subcritical* regime during the time of the delta variant.

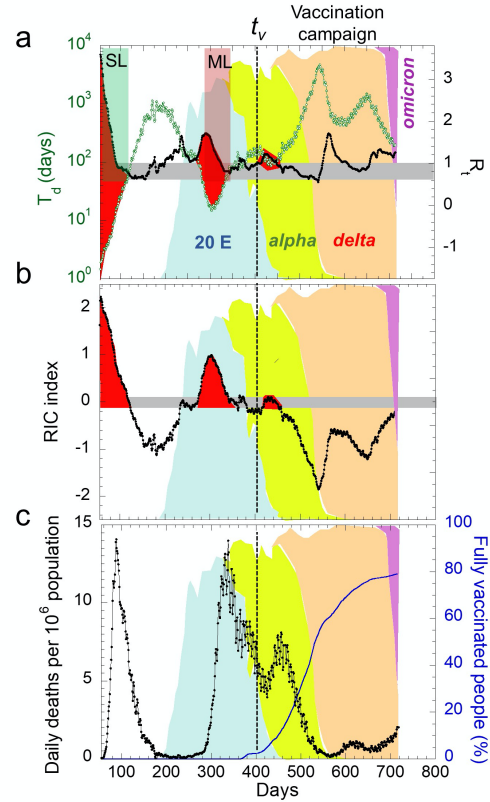

**Figure 1. Epidemic spreading in Italy from 24/02/2020 to 15/12/2021. (a)** Plots of the time dependent doubling time, $T_d$, (green dots) and reproductive rate $R_t$ (black dots) in Italy from 24/02/2020 to 15/12/2021. The gray rectangle represents the critical region separating the supercritical ($T_d<50$, $R_t>1$) from the subcritical ($T_d>100$, $R_t<1$). The supercritical phases correspond to the red areas. Before vaccination, the non-pharmaceutical control of epidemics has seen the application of strict lockdown (SL) followed by mild lockdown (ML) with loos rules. The intervals associated with the diffusion of the different coronavirus variants in Italy, European 20E, *alpha*, *delta* and *omicron* [38,39] are indicated. **(b)** *RIC-index* (black dots) as a function of time. The supercritical phases occur for positive values of the *RIC-index*. **(c)** Daily new deaths per million of population and percentage of fully vaccinated peoples (blue line) as a function time. We can distinguish the different waves, due to different coronavirus variants: the 1° and 2° waves are due to the *alpha* variant, the 3° is due to the *delta* variant. We note as the daily deaths number is suppressed by the vaccination campaign diffusion during the *delta* variant spreading, avoiding the exponential growth in the supercritical phase. Finally, we get signatures of the beginning of the wave due to the last *omicron* variant.

In order to take into account both $R_t$ and $T_d$ to describe the contagiousness of the pandemic complex dynamics, we have used the parameter introduced in [21]

$$RIC\text{-}index = \log_{10}(R_t/T_d) - 2 \tag{1}$$

The *RIC-index* value is near to zero in the critical phase, assumes negative or positive values in the *subcritical* or *supercritical* phases, respectively, as shown in **Figure 1b**. In the supercritical explosive supercritical regimes, the *RIC-index* shows a clear peak indicated by the red area. Alongside the variation of the contagiousness of the pandemic, the other important factor to keep in consideration is the danger and the severity of the disease.

At this aim we consider the number of daily fatalities per million population, *Df*, shown in **Figure 1c** where the peaks of the recurrent waves correspond with the peaks of the *RIC-index*. The figure shows also the percentage of fully vaccinated people versus time. It is clear how fatalities number is suppressed at larger percentage of vaccinated people, although further *recurrent waves* with different duration and intensity in the *subcritical* phase have occurred during the vaccination campaign. In particular, we note first the suppression of a third wave (around day 430, after $t_v$=400 day in Figure 1c) due to the diffusion of the *alpha* variant. In this period and afterwards, during the *delta* variant diffusion, the vaccination campaign lowers the number of deaths and prevents the spreading rate from falling into the *supercritical* phase, being the *RIC-index* below threshold.

In order to get quantitative information on the variation of recurrent traveling waves with the onset of vaccination campaign, we have applied the wavelet transform (WT) on both the *RIC-index* and the daily fatalities per million population time series, *Df*, and we have calculated the Largest Lyapunov Exponents map of the *RIC-index* versus *Df* in the time dependent orbit.

**2.2 Wavelet and Fourier transform of periodic recurrent waves**

We have applied the WT to the *RIC-index* and *Df* of Covid-19 pandemics evolution to visualize the different periodic fluctuations of Covid-19 spreading in magnitude scalogram color plots. WT are also useful for analysing weak and localized discontinuities, such as metastable phases [21,22]. The WT algorithm has been used to analyze the frequency structure of both the *RIC-index* and *Df* time series using the Morse wavelet (see Methods). The *RIC-index* scalogram, given by the local wavelet power spectrum, LWPS, in **Figure 2a** shows three main periods longer than 50 days indicated by the white dashed lines. Shorter periods are assigned to metastable phases.

The three main periods increase with the progress of the massive vaccination campaign after $t_v$=400 day which is also shown by the global wavelet power spectrum, GWPS plotted in **Figure 2b**.

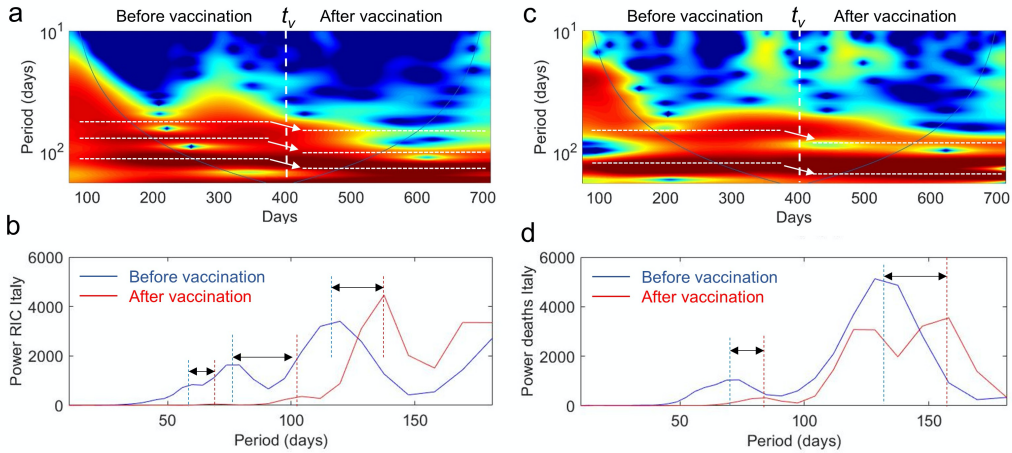

**Figure 2. Wavelet and Fourier Transform of Covid-19 periodic travelling waves in Italy from 24/02/2020 to 15/12/2021.** The Covid-19 spreading is visualized as a function of time by wavelet transform of **(a)** *RIC-index* and **(c)** daily deaths per million population, *Df*, time series. The wavelet analysis decomposes the spreading waves at different scales hidden in time series data. The main periods, before and after the vaccination campaign started at $t_v$=400 day, are indicated by the white horizontal dashed lines. We note a shift towards longer periods with the activation of the vaccination campaign in both *RIC-index* and *Df* time series. We can also visualize shorter lived transient events during the pandemic spreading, corresponding to lower intensity spots at smaller periods. Power is color coded from -2.5 (blue) to 1 (red). Global wavelet power spectrum (GWPS) of periods of **(b)** *RIC-index* and **(d)** daily new deaths time series calculated for $t < t_v$ (blue line) before the vaccination campaign and for $t > t_v$ (red line), after the vaccination campaign. The horizontal arrows represent the periods elongation obtained with the vaccination campaign.

The GWPS is obtained by averaging the LWPS across time (see Methods) before and after $t_v$, giving analogous information of the traditional Fourier spectrum, but in different time windows. The key result presented in this work is evidence of the shift of the periods before and after the vaccination campaign, as indicated by the arrows. The periods pass from 58, 75 and 118 days before the vaccination campaign to 68, 105 and 138 days, during the vaccination campaign.

Similar behavior can be seen in the LWPS and in the GWPS of the *Df* time series in **Figure 2c** and **2d**, respectively, where two main periods result elongated with the vaccination campaign. In particular, the first period passes from 65 to 85 days crossing $t_v$, while the second period of 135 days before vaccinations splits in two periods of 125 and 155 days during the vaccination campaign. The time evolution of the observed Covid-19 recurrent waves is in qualitative agreement with the results of the wavelet analysis applied to periodic travelling waves of measles in London [3].

**2.3 *Largest Lyapunov MAP* of *RIC-index* versus daily deaths helicoid time trajectories.**

The periodic *recurrent waves* can be visualized in a three dimensional space, probing the contagiousness of pandemic spreading correlated with fatality numbers as a function of time. By plotting the time evolution of the *RIC-index* and *Df* we obtain a non-linear helicoid vortex shown in **Figure 3a** composed by different acentric rings suggesting a chaotic dynamical evolution where the instabilities zones can be given by the Lyapunov exponents of dynamical evolving trajectories along the 3D orbit in Figure 3a.

The analysis of the helicoidal evolution of the time series of the *RIC-index* and the *Df* as function of time are available only over a short limited time interval and it is driven not only by the classical diffusion and random infection by contact events between humans in the macroscopic physical world, but also by the probability rate of the unpredictable virus gene point mutations because of unpredictable atoms replacements or deletions in the microscopic many body quantum world.

Since it is not possible to interpret our time series in term of classical Lyapunov exponents we have used the data-driven method by Rosenstein [43] to estimate the Largest Lyapunov Exponent (LLE) for short time series, used and tested in a wide variety of complex systems [44-48] like biological time series characterized by transient states of living matter which diverge from the basic assumptions for a classical chaotic system.

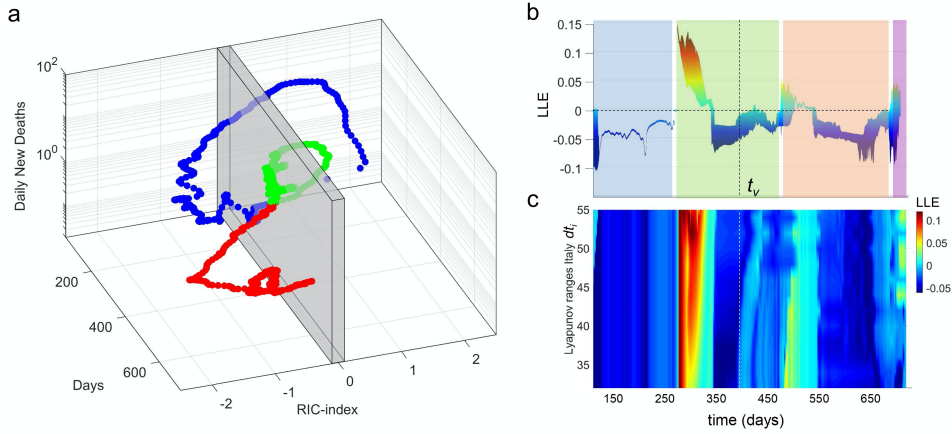

**Figure 3. Non-linear pandemic dynamics: *LLE MAP* of *RIC-index* versus daily deaths time trajectories. (a)** 3D orbit of the daily new deaths per millions of population and the *RIC-index* as a function of time (days) in Italy. The gray barrier represents the critical crossover between the *supercritical* and *subcritical* regimes. The different colors represent the three pandemic waves defined in the *Lyapunov Map* shown in panels **(b)** and **(c)**. The different waves can be individuated by the instability zones, where the LLE becomes positive and the 3D orbit diverges, indicated by the different colored regions. The first discontinuity point at day 280, corresponds to a quite strong instability explosion due to the second wave with the European *20E* mutation diffusion. The next discontinuity point occurs around day 470 in correspondence with the *alfa* mutation occurrence. Afterwards, during the *delta* variant diffusion, the LLE is kept on negative values thanks to the vaccination campaign that avoids new discontinuities towards positive LLE values. Finally, we find the third point of discontinuity with the beginning of the last wave associated with the new COVID 19 variant called *omicron*, around the day 700, where the LLE return positive.

In our approach we have chosen several initial ranges in the 3D orbit for LLE calculations. In order to follow the dynamics of the system day per day, we have considered all orbits on the intervals $\Delta t_i^0 = [t_i^0 \; t_i^0 + dt_i]$ days, where $t_i^0 = 73$ days is the initial day and $dt_i$ is an initial range spanning from $dt_1 = 32$ days up to $dt_{24}$ 56 days, which is the lowest pandemic period, obtained by WT analysis. We have calculated the Largest Lyapunov Exponent for all orbits obtained by increasing the interval $\Delta t_i^0$ by one day until the LLE becomes positive, at the $t_i^1$ day. This point, called *discontinuity point*, becomes the new initial instant, and the LLE is calculated on the new $\Delta t_i^1 \; [t_i^1 \; t_i^1 + dt_i]$ interval. This interval is increased by one day at each LLE calculation up to a new discontinuity point $t_i^2$. The procedure is iterated up the last point of the orbit (see Methods). Changing the initial range $dt_i$ with $1 < i < 24$, we obtain a *Lyapunov MAP*, shown in **Figure 3b** and **3c**. The LLE exponent in each point has been calculated using the Rosenstein approach which ensures accuracy and fast computation for short and noisy data sets [43]. We underline that a positive value of LLE indicates divergent trajectories and unstable system whilst a negative value indicates convergent orbits attracting towards a stable fixed point or periodic orbit. A positive LLE represents the instability zones where pandemic spreading results unstable and unpredictable. Than the LLE return to negative values in convergent trajectories with stable fluctuations to explode again in the next discontinuity point. We assume these discontinuity points as the limits of successive pandemic waves in the *RIC-index - Df* space, represented by different colored rings in the 3D trajectory of Figure 3a.

After the 1$^{st}$ wave due to the Wuhan variant (blue ring in Figure 3a and blue light rectangle in Figure 3b), we identify the explosion of the 2$^{nd}$ wave by the positive decreasing of the LLE from positive toward negative values, during the diffusion of the *European 20E* variant (green ring in Figure 3a and light green rectangle in Figure 3b). A 3$^{rd}$ wave due to the *alpha* variant diffusion can be identified by the fluctuations the LLE in the light green rectangle, soon after the beginning of the vaccinations campaign at $t_v$. In this case the fluctuation remains stable, since negative LLE does not exceed the zero values. After this, we observe a new discontinuity with increasing LLE towards positive values in correspondence with the *delta* variant spreading around day 470. This is followed by a stabilization of the orbit with negative LLE ascribed to the vaccination campaign which avoids the exponential growth (red ring in Figure 3a and light red rectangle in Figure 3b), as found in Figure 1. Finally, we get evidence of the start of the new omicron variant spreading by the new instability zone around day 700.

**METHODS**

*Wavelet analysis.* Traditional Fourier analysis partitions the total power (variance) of the time series between sinusoidal components at different frequencies. This approach lacks information on the time sequence of frequencies. By contrast, the wavelet approach allows to study recurrent waves phenomena in time as well as frequency. Rather than a sinusoid, the method is based on the Morse wavelet function [40-42], which capture local (in time) cyclical fluctuations in the time series. As with all wavelets, the frequency-time range over which it does this is set by a scale parameter, s. In general, wavelet scale is related to the conventional Fourier period of oscillations. Local wavelet power spectra, LWPS, of the *RIC-index* and daily deaths per millions of populations time series are presented in Figure 2a and 2c. Power is colour-coded as shown. Global wavelet power spectrum, GWPS, of both *RIC-index* and daily deaths per millions of populations time series have been calculated by averaging the LWPS across time, giving standard Fourier transform. Waves in the epidemic generates a peak in power spectra. The superimposed parabola represents the cone of influence, which measures the extent of edge effects. The WT has been performed, after smoothing accomplished using Gaussian windows on 14 days, for avoiding higher frequencies due to data collection artefacts during e.g. in the public holidays and weekends.

**Largest Lyapunov Exponent.** Classical Lyapunov exponents provide a measure of the exponential growth due to infinitesimal perturbations on a time series. A classical dynamical system becomes *chaotic* when very close starting states will diverge exponentially for some short time. The rate of exponential divergence is measured with classical Lyapunov exponents. If two initial states are separated by a small distance $s_0$, then for a short time interval the separation will evolve according to the equation $s(t) \approx s_0 e^{\lambda t}$. The value $\lambda$ is defined as the Largest Lyapunov Exponent, LLE, of the system. Chaotic systems show positive Lyapunov exponents and their sensitivity to initial conditions increases with $\lambda$. Nearby points of an orbit will diverge to any arbitrary separation; therefore, a larger the exponent shows a more unstable system. On the contrary, a negative Lyapunov exponent indicates that the orbit converges to a stable fixed point, as generally occurs in a dissipative system with asymptotic stability; the more negative the exponent, the greater the stability. Finally, the points in the orbit where $\lambda=0$, are fixed points where the system is in a steady-state mode or near the transition to chaos. The extraction of the classical Lyapunov exponent assumes that the system follows deterministic trajectories over an unlimited time period over a very long time with no transient states, typical of biological systems and the data analysis requires very long time series which are typically not available for epidemics spreading.

Therefore for the Covid-19 short time series we have calculated the Largest Lyapunov Exponents following the Rosenstein algorithm [43]. In this approach it is first necessary to reconstruct the state space from the experimental data record. The original time series data and its time-delayed copies determine the topological structure of a dynamical system [50]:

$$Y = [X(t), X(t + T), ..., X(t + (d_E - 1)T)] \qquad (3)$$

where $Y$ is the reconstructed $d$-dimensional state vector, $X(t)$ is the observed variable, $T$ is a time lag, and $d_E$ is the embedding dimension. This method looks for the nearest neighbor of each point in phase-space and tracks their separation over a certain time evolution. This separation is also a function of the location of the initial value. To obtain LLE on different initial times, we have built a Lyapunov map. Each point in this map corresponds to a Lyapunov exponent, $L\{Y_i^j(t)\}$, calculated for the time series, $\{Y_i^j(t)\}$, starting from an initial instant, $t_i^j$, up to an initial ranges, $dt_i$, and increasing by one the successive series up to the discontinuity point, $t_i^{j+1}$, where the Lyapunov exponent goes from a negative value to a positive value. These points, $t_i^j$, where the convergent trajectory becomes divergent, are called *discontinuity points* and indicate the transition between stable and unstable regimes. As soon as the Lyapunov exponents become positive, we identify the *discontinuity points* as initial instants $t_i^{j+1}$ of the new series $\{Y_i^{j+1}(t)\}$. The procedure is iterated for the $\{Y_i^{nj}(t)\}$ time series corresponding to the $t_i^{nj}$ point of discontinuity. The index, $i$, represents a given initial range chosen from a minimum (32 days) to a maximum value of 56 days not exceeding the minimum non-local period in the wavelet transform calculation.

Thus a Lyapunov exponent of a map with pixels $(i,j)$ can be written as

$$L\{Y_i^j(t)\} = L\{Y[Ric(\Delta t_i^j), Df(\Delta t_i^j)]\} \qquad (4)$$

where

$$\Delta t_i^j = (t_i^j, ..., t_i^i + dt_i, t_i^j + dt_i + 1, ..., t_i^j) \qquad (5)$$

*Ric* and *Df* are the *RIC-index* and the daily deaths per million of population. In our case a Lyapunov map is made of 24 horizontal profiles, ($1<i<24$); each profile is calculated using the initial range, $dt_i=31+i$. Each horizontal profile is characterized by $1<j<n_i$ discontinuity points where $L\{Y_i^j(t)\}$ change sign becoming positive. This procedure allows us to visualize and identify the *discontinuity points* that are independent from the assumed initial ranges, $dt_i$, in calculations. Quantitative determination of the discontinuity points on the helicoid 3D trajectory is made by finding the discontinuity points in the averaged profile of LLE, in the map. As for the WT analysis, the Lyapunov Map is calculated on time series smoothed on Gaussian windows of 14 days, for avoiding higher frequencies due to the mentioned data collection possible artefacts.

## 3. Conclusions

We have studied the periodic recurrent waves of the pandemic COVID-19 by analyzing its contagiousness (*RIC-index*) and the impact on country population (the daily deaths per million of population $D_f$) as a function of time in a 3D space parameter. The wavelet transform approach shows that the pandemic periods are elongated during the vaccination campaign. As a result, we present a chaotic helicoid-like vortex behavior. Our unconvential chaotic map is obtained by calculating the time dependent Largest Lyapunov Exponents of a reconstructed space made of 3D orbits evolving day per day of fatalities number versus *RIC-index*. The detection of any possible crossover from negative to positive LLE gives the instability zones corresponding to the onset of new pandemic waves. We have shown that the non-linear time evolution of the fatalities number and the *RIC-index* in the chaotic vortex provides a quantitative tool needed for prompt response to appearing new waves due to a new mutant of the coronavirus.

In conclusion, we have found that vaccination increases the period of the oscillations, which is crucial for the control action. However, further work is in progress to monitor discontinuities points occurring after the vaccination related with the appearing of the new variants and duration of effectiveness of covid vaccine. The use of the proposed approach in the containment policies will provide early warning to critical situations and quantitative measure of the slowing down the rate of diffusion of the virus during the vaccination campaign.

## Acknowledgments

This work was funded by Superstripes onlus.

**Author Contributions:** The authors equally contributed to this work.

**Competing Interest Statement:** Authors declare that they have no competing interests